\documentclass[prl,twocolumn,superscriptaddress,showpacs]{revtex4}
\usepackage{graphicx}
\newcommand{\be}{\begin{equation}}
\newcommand{\ee}{  \end{equation}}
\newcommand{\ba}{\begin{eqnarray}}
\newcommand{\ea}{  \end{eqnarray}}

\begin{document}
\title{Kondo temperature for a quantum dot in an Aharanov--Bohm ring}
\author{C. H. Lewenkopf}
\affiliation{Instituto de F\'{\i}sica, UERJ,  
             R. S\~ao Francisco Xavier 524, 20550-900 Rio de Janeiro,
             Brazil}
\author{H. A. Weidenm\"uller}
\affiliation{Max-Planck-Institut f\"ur Kernphysik, D-69029
Heidelberg, Germany}

%%%%%%%%%%%%%%%%%%%%%%%%%%%%%%%%%%%%%%%%%%%%%%%%%%%%%%%%%%%%%%%%%%%%%%%%%%
%                            ABSTRACT                                    %
%%%%%%%%%%%%%%%%%%%%%%%%%%%%%%%%%%%%%%%%%%%%%%%%%%%%%%%%%%%%%%%%%%%%%%%%%%

\begin{abstract}
We study the Kondo temperature of a quantum dot embedded into one arm
of an Aharonov--Bohm interferometer. The topology of a disordered or
chaotic Aharanov--Bohm ring leads to a stochastic term in the scaling
equation and in the renormalization procedure. As a result, the Kondo 
temperature displays significant fluctuations as a function of magnetic
flux.
\end{abstract}

%
% 03.65.Yz  Decoherence; open systems; quantum statistical methods 
% 72.10.-d  Theory of electronic transport; scattering mechanisms
% 72.10.Bg  General formulation of transport theory
% 72.10.Fk  Scattering by point defects, dislocations, surfaces, and
% other imperfections (including Kondo effect)
% 72.15.Qm  Scattering mechanisms and Kondo effect [electronic
% cond. in metals]
% 72.20.Dp  General theory, scattering mechanisms [conductivity in
% semiconductors]
% 73.21.-b  Electron states and collective excitations in multilayers,
%           quantum wells, mesoscopic, and nanoscale systems
% 73.21.La  Quantum dots
% 73.23.Hk  Coulomb blockade; single-electron tunneling 
% 73.23.-b  Mesoscopic systems 
% 73.23.Ad  Ballistic transport 
% 73.63.-b  Electronic transport in mesoscopic or nanoscale materials
% and structures
% 73.63.Kv  Quantum dots

\maketitle

Recently, the transmission phase of electrons passing through a
quantum dot (QD) embedded in an Aharanov--Bohm (AB) interferometer has
been measured~\cite{Ji00}. This was done in the Kondo regime and
over an energy interval given by the width of the ``odd'' valley
separating two conductance Coulomb--blockade peaks. The QD with
half--integer spin plays the role of the Kondo
impurity~\cite{Glazman88}. Theory predicts the phase shift to be 
$\pi$ at low temperature~\cite{Gerland00}. The results are
puzzling and do not follow this prediction. In the present Letter we
show that the $T$-matrix ${\cal T}_{\rm K}$ of the Kondo problem
acquires specific properties due to the topology of the AB ring.
These have not been taken into account previously and profoundly
affect the dependence of ${\cal T}_{\rm K}$ on the AB phase. They
serve as one example for the role of finite--size
effects~\cite{Simon02} and mesoscopic fluctuations~\cite{gla} which,
together with experiments on semiconductor quantum
dots~\cite{Goldhaber-Gordon98} and nanotubes~\cite{Nygard00} have
rekindled interest in the Kondo effect~\cite{hews} recently.
Implications of our results for the experimental analysis are briefly
discussed.

{\it Model.} 
We consider a QD carrying an odd number of electrons in the Kondo
regime. The QD is embedded into one arm of an AB ring
threaded by a weak magnetic flux $\Phi$. We take account only of the
AB phase $\phi = 2 \pi \Phi / \Phi_0$. Here $\Phi_0$ is the elementary
flux quantum. The AB ring is attached to $\Lambda$ leads labelled $j =
1, \ldots, \Lambda$. Without counting spin degrees of freedom labelled
$s = \pm 1/2$, lead $j$ carries $N_j$ channels (transverse modes with
energies below the Fermi surface) labelled $a = 1, \ldots, N_j$. The
total number of channels is $N = \sum_j N_j$.  

The leads are separated from the interior of the AB ring by fictitious
barriers, and the interior from the QD by real tunneling barriers
caused by the gate potentials. With $E$ the energy, the lead
Hamiltonian is $H_{\rm L}$ $= \sum_{j=1}^{\Lambda} H_j$ $= \sum_j
\sum_{a, s} \int {\rm d}E \; E \; c^{\dagger}_{j a s E} c^{}_{j a s
E}$, with $c^{\dagger}$ and $c$ the creation and annihilation
operators obeying $\{c^{\dagger}_{j a s E}, c^{}_{j' a' s' E'} \} =
\delta_{j j'} \delta_{a  a'} \delta_{s s '} \delta(E - E')$. The
Hamiltonian $H_{\rm AB}$ of the AB ring has a discrete spectrum with
eigenvalues $\epsilon_{\mu}$ with $\mu = 1, 2, \ldots$ and is given by
$H_{\rm AB}$ $= \sum_{\mu s} \epsilon_{\mu}$ $d^{\dagger}_{\mu s}
d^{}_{\mu s}$ where $\{ d^{\dagger}_{\mu s}, d^{}_{\nu s'} \} =
\delta_{s s'} \delta_{\mu  \nu}$. Leads and AB ring are coupled by the
tunneling Hamiltonian $H_{\rm AB-L}$ $= \sum_{s j} \sum_{\mu} \int
{\rm d}E \ W_{j a; \mu}(E) \ [ c^{\dagger}_{j a s E} d^{}_{\mu s} +
{\rm H.c.} ]$. We assume that the real tunneling matrix elements $W_{j
  a; \mu}(E)$ change little over an energy scale given by the width of
the Kondo resonance. We also assume that the levels $\epsilon_\mu$ lie
so dense (and that their coupling to the leads is so large) that all
scattering through the AB ring in the absence of the QD is smooth in
energy on the same scale.

The tunneling matrix elements $v^P_{\mu}$ between the QD and state
$\mu$ of the AB ring carry an upper index $P = L,R$ denoting tunneling
through the left (right) barrier, respectively. The occurrence of two
amplitudes for each state $\mu$ of the AB ring reflects the topology
of the ring. We choose the matrix elements $v^P_{\mu}$ real and
display the AB phase $\phi$ explicitly in the Kondo Hamiltonian. Gauge
invariance allows us to put the entire effect of the AB phase $\phi$
onto the right barrier separating the QD from the AB ring. Whenever
the electron leaves (enters) the QD through that barrier, it picks up
the phase factor $e^{i \phi}$ ($e^{- i \phi}$, respectively). We
define $v_\mu = v^L_\mu + e^{i \phi} v^R_\mu$.

We focus attention on the Coulomb blockade mid--valley region. Then it
is justified to use the s--d model for the Kondo resonance~\cite{hews}. 
(We have convinced ourselves that our arguments hold likewise for the
Anderson model~\cite{hews}. However, we have not yet performed a
complete analysis of the latter model). In the s--d model, a spin ${\vec
  S}$ (which represents the total spin of the electrons on the QD) is
coupled to the spins of the electrons in the AB ring,
\be
H_{\rm K} = \frac{2}{E_C} \sum_{{\mu s}\atop{\nu s^\prime}} v_{\mu}
d^{\dagger}_{\mu s} ({\vec \sigma}_{s s'} \cdot {\vec S}) v^*_{\nu} d^{}_{\nu
  s'} \ .
\label{4}
\ee
Here $E_C$ is the charging energy of the QD, and ${\vec \sigma}$
stands for the vector of the three Pauli spin matrices. The total
Hamiltonian is the sum of these terms,
\be
{\cal H} = H_{\rm L} + H_{\rm AB} + H_{\rm AB-L} + H_{\rm K} \ . 
\label{6}
\ee

Using standard techniques~\cite{ramm,gla} we obtain for the
non--diagonal dimensionless conductance coefficients $g_{jl}$ with $j
\neq l$ of the  many--body Hamiltonian ${\cal H}$
\be
g_{j l} = \!\int\!{\rm d}E \!\left(\! -\frac{{\rm
    d}f}{{\rm d}E} \right) \!\mbox{Tr}_{\mu s} \Big[ \Gamma_j(E) 
G^r(E)  \Gamma_l(E)  G^a(E) \Big] .
\label{eq:conductance}
\ee
Here $f$ is the Fermi function, $G^r$ and $G^a$ are the ordinary
retarded and advanced equilibrium many--body Green's functions of   
${\cal H}$, and $[\Gamma_j]_{\mu\nu} = 2\pi \sum_a W_{ja;\mu}
W_{ja;\nu}$.

In the absence of the Kondo term, we deal with a pure single--particle 
problem, with a Green's function $G^{r,a}_{0; \mu \nu}$ given by
$[G^r_{0}]^{-1}_ {\mu \nu} = (E - \epsilon_\mu) \delta_{\mu \nu} +
(i/2) \sum_j [\Gamma_j]_{\mu \nu}$, and non--diagonal elements of the
spin--independent scattering matrix $S^{(0)}_{ja; lb}(E) = - 2 i \pi
\sum_{\mu \nu} W_{j a; \mu} [G_0^{r}]_{\mu \nu} W_{l b;
  \nu}$~\cite{hack,haw}. Hence, as expected, the conductance is given
by the Landauer--B\"uttiker formula.

We return to the many--body case and expand $G^r_{\mu \nu}(E)$ in
powers of $H_{\rm K}$ to write $G^r_{\mu \nu}(E) = G^r_{0; \mu \nu}(E)
+ \big[ G^{r}_{0}(E) \ {\cal T}_{\rm K} G^{r}_{0}(E) \big]_{\mu \nu}$,
where 
\be
{\cal T}_{\rm K} = H_{\rm K} \sum_{n = 0}^\infty [ G^r_{0}(E)
H_{\rm K} ]^n
\label{38a}
\ee
defines the $T$--matrix for the Kondo effect. Using ${\cal T}_{\rm K}$
and the definition of $S^{(0)}$, we write $g_{j l}$ as
\ba
&&g_{j l} = \int
%_{-\infty}^{\infty}
\! {\rm d}E \ \left( -
\frac{{\rm d}f}{{\rm d}E} \right) \ \sum_{a b} \Bigg\{  2  \left| S^{(0)
}_{ja; lb}(E)  \right|^2 +\nonumber \\
&& 
\!\!\!\!\!\!\!\!\Big( (S^{(0)}_{ja; lb})^* \mbox{Tr}_{s}\! \left[ \!- 2
i \pi \sum_{\mu \nu} \ W_{j a; \mu} [ G^{r}_{0} \ {\cal T}_{\rm K} \
G^{r}_{0} ]_{\mu \nu} W_{l b; \nu} \right] + {\rm c.c.} \Big)
\nonumber \\ && 
\!\!\!\!\!\!\!\!+ \mbox{Tr}_{s} \Big[ \big| \!\!- 2 i \pi  \sum_{\mu
  \nu} \ W_{j a; \mu}(E) [ G^{r}_{0} \ {\cal T}_{\rm K} \ G^{r}_{0}
]_{\mu \nu} W_{l b; \nu} \big|^2 \Big] \Bigg\} \ .
\label{39}
\ea
Three processes contribute to transport through the AB ring, each
represented by one term on the r.h.s.\ of Eq.~(\ref{39}): Scattering
of an electron through the AB ring without its passing through the QD,
the interference term between the amplitude for that process and the
amplitude for (multiple) visits of the QD, and the square of the
latter amplitude. This structure differs fundamentally from that for a
QD in the Kondo regime connected to two leads without the topology
of the AB ring. Here only the interference term occurs~\cite{gla}.
This is caused by the special form of the coupling between Kondo
resonance and channels.

{\it Dependence upon the AB Phase.}
We compare the dependence of $g_{j l}$ on the AB phase $\phi$ with
that obtained for a simpler problem: An AB ring with a QD that 
carries a single level at energy $E_0$ and causes a Breit--Wigner
resonance~\cite{haw}. The resulting expression for  $g_{j l}$ 
($j \neq l)$ has the form of our Eq.~(\ref{39}) except for the
replacement~\cite{haw} $- 2 i \pi W_{j a; \mu}(E)$ $[ G^{r}_{0}(E)$
${\cal T}_{\rm K}$ $G^{r}_{0}(E) ]_{\mu \nu}$ $W_{l b; \nu}$ $\to$
$- i ( \gamma_{j a} + \eta_{j a} e^{i \phi} )$ $(E - E_0 + i
\Gamma/2)^{-1}$ $( \gamma_{l b} + \eta_{l b} e^{- i \phi} )$. The
amplitudes $\gamma_{j a}$ and $\eta_{j a}$ are complex and independent
of energy $E$ and AB phase $\phi$. The only dependence of the
resulting expression on the phase $\phi$ is that given explicitly and
that of the total width $\Gamma = \sum_{j a} | \gamma_{j a} + e^{i
  \phi} \eta_{j a} |^2$. Under the assumption that the electrons
move chaotically and/or diffusively in the AB ring, the dependence of
$\Gamma$ on $\phi$ is~\cite{haw} of order $1/ \sqrt{N}$ and, thus,
negligible whenever $N \gg 1$. The experimental determination of
the transmission phase through the QD (defined as the phase of the
Breit--Wigner term) is, thus, possible for $N \gg 1$.

In the case of Eq.~(\ref{39}), the $T$--matrix ${\cal T}_{\rm K}$
depends non--trivially on $\phi$, see Eqs.~(\ref{38a}) and (\ref{4}).
Thus, the central question is: How strong is this dependence of 
${\cal T}_{\rm K}$ on $\phi$? The theoretical treatment of the Kondo
resonance requires scaling and/or renormalization techniques. Our
question can only be answered in this framework.

{\it Poor Man's Scaling.} The essential aspects of poor man's scaling
emerge already when one considers the matrix element of the
$T$--matrix to second order in $H_{\rm K}$, given by $\langle \nu
s^\prime$ $| H_{\rm K} G^r_0(E)$ $H_{\rm K} | \mu s \rangle$. We
suppress the external factors $v_\mu$ and focus on the
intermediate--state summation involving $G^r_0(E)$. Two processes
contribute to this term: The scattering of an electron and that of a
hole. Aside from spin--dependent factors, the first process yields
\ba
\label{46}
&&\sum_{\tau \rho} v_\tau G^{r}_{0; \tau \rho}(E) v^*_\rho =
\sum_{\tau} \frac{|v_\tau|^2}{E^+ - \epsilon_\tau}  \\
&& \!\!\!\!\!\!\!\!\!\!\!\!
+ \sum_{\tau \rho} v_\tau \frac{1}{E^+ - \epsilon_\tau} \Big[\!\! - i
\sum_{j a; l b} W_{\tau; j a} F_{j a;l b} W_{l b;\rho} \Big]
\frac{1}{E^+ - \epsilon_\rho} v^*_\rho \nonumber
\ea
where $[F^{-1}]_{j a ; l b}$ $= \delta_{j l} \delta_{a b}$ $+ i
\sum_\mu W_{j a; \mu}$ $[ E^+ - \epsilon_\mu]^{-1} W_{\mu;l b}$.  
A similar decomposition into a term involving only the states in the
AB ring and a remainder applies to hole scattering. Our problem
differs from the standard Kondo situation of a magnetic impurity
coupled to the conduction band~\cite{hews} in three main respects, the
last two being related to the topology of the AB ring: (i) The
energies $\epsilon_\tau$ do not lie in the conduction band but are
defined as the eigenvalues of electrons moving in the AB ring. Thus,
the summation over $\tau$ extends to $+\infty$. (ii) $G^r_0(E)$
differs from the Green's function for free electrons. This leads to
the $2^{\rm nd}$ term on the r.h.s.\ of Eq.~(\ref{46}). (iii) The
$v_{\mu}$'s involve the AB phase $\phi$. We address these in turn.

(i) The coupling matrix elements $v^L_\tau$ and $v^R_\tau$ reflect the
dynamics of the coupling of the QD to the AB ring. The $v_\tau$'s are
overlap integrals involving the wave function $\chi$ of the electron
on the QD and that of states $\tau$ with mean level spacing $\Delta$
in the AB ring. Two tendencies characterize the behavior of the
$v_\tau$'s. (a) As the number of nodes of the state $\tau$ in the
overlap region becomes much bigger or smaller than that of $\chi$, the
overlap is strongly reduced. (b) In a diffusive and/or chaotic AB
ring, the states $\tau$ within an energy interval given by the
Thouless energy $E_T$, are strongly mixed. We take account of both
features in terms of a cutoff model. Within an energy window of
bandwidth $D \leq E_T$ around the energy of the state $\chi$, the
$v_\tau$'s are~\cite{haw} uncorrelated Gaussian distributed random
variables with mean value zero and a common constant variance $v^2$
while the $v_{\tau}$'s for states $\tau$ outside of this window
vanish. For both ballistic and diffusive rings $E_T$ is large compared
to the mean level spacing $\Delta$ of the states in the AB ring but
small in comparison to the Fermi energy. The cutoff then applies
likewise to the summation over particles and to that over holes. The
cutoff restricts the summation over $\tau$ to a finite number $g
\approx E_T / \Delta$ states, with $g \gg 1$. To identify the
influence of the large contributions to Eq.~(\ref{46}) and to resum
them, we let $g \rightarrow \infty$ and use scaling.

(ii) Scaling needs to be applied only to the first term on the
right--hand side of Eq.~(\ref{46}) since only this term produces a
logarithmic singularity at the upper band edge. As for the second
term, the expression in brackets is obviously finite. There are two
contributions to the summations over $\tau$ and $\rho$. For $\tau =
\rho$, we sum squares of amplitudes $v_\tau$ but also of the
propagator. The result is inversely proportional to $D$ and vanishes
for $D \to \infty$. For $\tau \neq \rho$, we use that the $v_\mu$'s
are Gaussian random variables with mean value zero. Their random signs
quench the intermediate--state summations and ensure that the result
is convergent even as $g \to \infty$. Thus, the singular contributions
to the second--order particle matrix element are obtained by replacing
$G^r_0(E)$ by $[ E^+ - H_{\rm AB}]^{-1}$. Corresponding statements
apply to the terms of higher order in $H_{\rm K}$. These arguments
apply likewise to hole scattering. Thus we have reduced our problem to
one which is quite similar in appearance to the standard Kondo case:
Singular terms arise only from the coupling of the QD to the states in
the AB ring, i.e., from the Hamiltonian $H_{\rm AB} + H_{\rm K}$. The
terms $H_{\rm L}$ and $H_{\rm AB-L}$ in ${\cal H}$ of Eq.~(\ref{6})
are irrelevant for scaling.

(iii) We turn to the AB phase appearing in the intermediate--state
summation. The first term on the right--hand side of Eq.~(\ref{46})
contains the phase through $|v_\tau|^2$ $= (v^L_\tau)^2$ $+
(v^R_\tau)^2$ $+ 2 \cos \phi$ $\ v^L_\tau v^R_\tau$. To define the
scaling variable, we write this expression in the form
$[(v^{L}_\tau)^2$ $+ (v^{R}_\tau)^2]$$[1 + \cos \phi \cos \psi_\tau]$
where
\begin{equation}
\cos \psi_\tau = \frac{2 v^L_\tau v^{R}_\tau}{(v^{L}_\tau)^2 +
  (v^{R}_\tau)^2} \ . 
\label{51}
\end{equation}
Since the $v_{\tau}$'s are uncorrelated Gaussian variables,
$\psi_\tau$ is also a random variable which is uniformly distributed
in the interval $0 \leq \psi_\tau \leq \pi$. Moreover, the
$\psi_\tau$'s with different indices $\rho$ are uncorrelated. We
replace the positive quantity $[(v^{L}_\tau)^2 + (v^{R}_\tau)^2]$ by
its average $2 v^2$. An equivalent assumption (independence of the
coupling strength on the index $\tau$ characterizing states in the
conduction band) is usually made in the standard Kondo case. We
checked that this assumption does not affect the generic features of
our results.

We define the scaling variable $J = 4 v^2 / E_C$ and apply the standard 
algebra of poor man's scaling~\cite{hews} to write
\begin{equation}
\frac{\delta J}{\delta D} = - \frac{2 \rho_0 J^2}{D} (1 + \cos \phi \cos
\psi) \ .
\label{52}
\end{equation}

Here $\rho_0 = \Delta^{-1}$ is the density of states $|\mu \rangle$ in
the AB interferometer. We assume that $\rho_0$ is constant. Because of
the appearance of the random variable $\psi$, Eq.~(\ref{52}) is a
stochastic differential equation. To solve it, we first consider the
case without stochastic term and define the ``standard scaling
trajectory'' by choosing a Kondo temperature $T_K$ and initial values
$\rho_0 D_0$ (where $D_0 \approx E_T$) for the bandwidth and $\rho_0
J_0$ for the coupling strength such that the relation $T_K = D_0 \exp(
- 1/2 \rho_0 J_0)$ holds. We then solve Eq.~(\ref{52}) by reducing
$\rho_0 D$ in unit steps drawing $\psi$ from a random--number
generator. We thereby eliminate the states of the AB ring one by one.
In the $p^{\rm th}$ step, the coupling constant $\rho_0 J$ is
``kicked'' out of one scaling trajectory and into another by the term
proportional to $\cos\phi \cos \psi_p$. The effect becomes larger as
$\rho_0 D$ is decreased and $\delta J / \delta D$ increases. As a
result, every realization $\{\psi_p \}$ of the sequence of
$\psi$--values determines a different scaling trajectory given by a
discrete sequence of points $\{ D_p, J_p \}$. The scaling process has
necessarily to end when all ring states are reduced to a single
remaining one that absorbs all interaction effects, giving a lower
energy cutoff $\Delta$. The lower cutoff is modified for two reasons:
(i) The AB ring is open, and the states $\mu$ have finite widths with
mean value $\Gamma = \Delta N / (2 \pi)$. We have neglected $\Gamma$
by omitting the second term on the r.h.s. of Eq.~(\ref{46}). This is 
justified only as long as $D \gg \Gamma$. (ii) The electrons which
screen the QD spin, form a Kondo cloud of size $\xi$ in the AB ring.
The energy scale $E_\xi$ which is required to attain a state of
correlation length $\xi$, can be estimated using the uncertainty
relation. Hence, the lower cutoff is $\lambda = \max\{\Delta,
\Gamma, E_\xi\}$.

We have solved the discretized version of Eq.~(\ref{52}) for a choice
of parameters which resembles the weak Kondo coupling experiment in
Ref.~\cite{Ji00}: (i) $\Gamma / \Delta \approx 10$ ($\Lambda = 6,
N_j\approx 10$); $E_\xi/\Delta \approx 1$ (since $\xi \approx 1 \mu$m
is about the size $L$ of the ring). (ii) $E_T / \Delta = \sqrt{4 \pi
  \,N_{\rm e}}$ where $N_{\rm e}$ is the number of electrons on the
ring. This formula follows for ballistic system from $E_T \approx h/L$
and the Weyl formula. The same formula yields $N_{\rm e} \approx 1000$.
(iii) $T_K/\Delta \approx 5$.

\begin{figure}[tbp]
%\vspace{0.7cm}
%\centering \leavevmode
%\center{\epsfig{file=sigmag.eps, ,width=8.0cm,angle=0}}
\includegraphics[width=8.5cm]{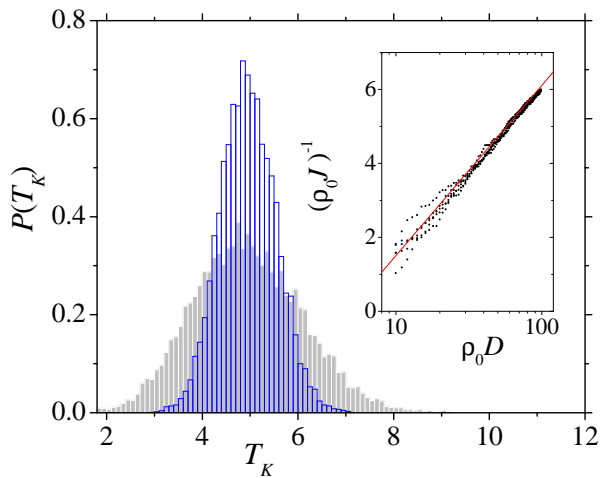}
\caption{Distribution of Kondo temperatures $T_K$, for $\cos\phi=0.5$
  (narrow distribution) and $\cos\phi = 1$ (wide distribution), both
  for $\langle T_K \rangle/\Delta = 5$. Each histogram contains $10^4$
  scaling trajectories. Insert: Typical scaling trajectories for 
  $\cos\phi=1$.}
\label{fig:P(T_K)}
\end{figure}

Our results are shown in Figure~\ref{fig:P(T_K)}. The insert shows
five representative trajectories for $\cos \phi = 1$ (where the
stochastic effect is largest). The deviations from the standard
scaling trajectory increase with decreasing bandwidth, as expected.
Averaging over the stochastic trajectories, we recover the standard
scaling trajectory (straight line in the insert). In the main part of
the figure, we show the distributions $P(T_K)$ of Kondo temperatures
$T_K$ for $\cos \phi = 1/2$ and $\cos \phi = 1$. These are obtained
from $10^4$ realizations of $\{ \psi_p \}$. For each realization, the
final value of $T_K$ is found by a best--fit procedure. Mean value
$\langle T_K \rangle$ and dispersion $\delta T_K$ are deduced from
$P(T_K)$. The average value $\langle T_K \rangle$ is very close to the
standard scaling value and does not show a significant dependence on
$\cos\phi$. The fluctuations are large, however, and $\delta T_K$
increases with $\cos\phi$, reaching a maximum value of $\delta T_K /
\langle T_K \rangle \approx 0.25$ at $\cos\phi=1$.

The numerical simulations show that the fluctuations become larger
(smaller) as either $\rho_0 T_K$ or the lower cutoff $\lambda$ are
decreased (increased). The reason is that the stochastic ``kicks"
become more effective for smaller values of $T_K$ and/or $D$, see
Eq.~(\ref{52}). Our results are less sensitive to $E_T$. As $E_T$ is
increased, the slopes of the scaling trajectories become dominated by
the upper band edge, were the stochastic kicks have little effect.
However, this happens very slowly: We observe only small quantitative
changes as we increase $E_T$ by an order of magnitude.

For the physical interpretation of our result, we say that an electron
is on the QD when it interacts with the fixed spin ${\vec S}$ of the
Kondo Hamiltonian $H_{\rm K}$. Each term of the perturbation expansion
describes a process or a sequence of processes where an electron
leaves the QD, moves in the AB ring, and re--enters the QD. Whenever
the electron leaves and enters the QD through the same barrier, the
result is a factor $(v^R)^2$ or $(v^L)^2$, as the case may be.
Processes of this type are not different from the ones that are taken
into account in the standard approach and lead to the standard form of
the scaling equation. The fact that the QD is coupled to the AB ring
via two barriers only contributes a factor two and is otherwise
irrelevant. There are, however, additional processes where the
electron leaves the QD through the left barrier and re--enters it
through the right one, or vice versa. Such processes correspond to a
complete loop through the AB ring, carry the factor $v^R v^L$ and the
AB phase, and are absent in the standard approach. Loops in either
direction are equally likely, hence the dependence of the relevant
term on $\cos \phi \ v^L v^R$. The AB ring is assumed to be chaotic or
diffusive. This makes the loop terms in the scaling equation
stochastic.

{\it Renormalization.}
Poor man's scaling breaks down for temperatures below the Kondo
temperature $T_{\rm K}$. We discuss this temperature regime only
qualitatively in order to demonstrate the existence and action of the
stochastic term. We use renormalization group arguments without
actually performing the calculation. Following standard
procedure~\cite{hews}, one would construct a new basis of orthonormal
states $| n \rangle$ where $n = 0, 1, 2, \ldots$. With $| {\rm vac}
\rangle$ denoting the vacuum, the state $| 0 \rangle$ is defined as
$|0\rangle = (\sum_\nu |v_\nu|^2)^{-1} \sum_\mu v_\mu^*d_\mu^\dagger |
{\rm vac} \rangle$. For $n \geq 1$, the states $|n \rangle$ would be
obtained with the help of Schmidt's orthogonalization procedure from
the sequence $H_{\rm AB} |0 \rangle$, $H_{\rm AB}^2 |0\rangle$,
$\ldots$. (We have shown that it is again justified to approximate the
Hamiltonian ${\cal H} - H_{\rm K}$ by $H_{\rm AB}$). The terms in the
resulting tridiagonal Hamiltonian~\cite{hews} depend upon the matrix
elements $v_\mu$ via expressions of the type $\sum_\mu v^2_\mu
(\epsilon_\mu)^m$, with $m = 0, 1, 2, \ldots$. Just as for poor man's
scaling, the averages (fluctuations) of such expressions are
independent of (dependent on) the AB phase. The fluctuations are
maximal for $\cos \phi = 1$ and vanish for $\cos \phi = 0$. Again, the
stochastic part is expected to cause a spread in the renormalization
group trajectories. The trend of the spread with temperature $T$ can
be ascertained as follows. The Kondo temperature $T_{\rm K}$ defines a
marginal fixed point of the problem, while $T = 0$ is a stable fixed
point. All trajectories converge to that point. Therefore, as the
sample is cooled down, the stochastic spread of the trajectories first
increases monotonically to reach its maximum at $T = T_{\rm
  K}$. Thereafter, the spread decreases monotonically and vanishes at
$T = 0$.

{\it Discussion.}
Because of the topology of the AB ring, the Kondo temperature $T_{\rm
  K}$ of a QD embedded into the ring depends stochastically on the AB
phase $\phi$. The resulting spread in Kondo temperature versus AB
phase is significant, amounting to 20 or even 40 percent. The
influence of stochasticity is expected to decrease as the temperature
is lowered below $T_{\rm K}$, and to vanish at $T = 0$. The precise
form of the dependence of $T_{\rm K}$ on $\phi$ cannot be predicted
for an individual experimental sample. We expect that for $T \approx
T_{\rm K}$, this stochastic dependence seriously limits attempts to
determine the transmission phase through the QD experimentally,
although quantitative predictions will have to be based upon a study of
the Anderson model. Such work is now in progress.

This work was partly done while one of us (HAW) visited Brazil under the
auspices of a CAPES -- Humboldt award. CHL is supported by CNPq and Instituto
do Mil\^enio de Nanoci\^encias. HAW
acknowledges stimulating discussions with Y. Gefen, Y. Imry, and Y. Oreg.

%---------------------------------------------------------------------------

\end{document}